\documentclass[prd,showkeys,floatfix,twocolumn,amsmath,amssymb,floatfix]{revtex4}
\usepackage{graphicx}
\usepackage{epstopdf}
\usepackage{multirow}
\usepackage{subfigure}
\usepackage{color}
\usepackage[colorlinks,citecolor=blue,anchorcolor=red,menucolor=red,linkcolor=red,filecolor=red,runcolor=red,urlcolor=blue,frenchlinks=red]{hyperref}
\usepackage{enumitem}
\usepackage{CJK}
\usepackage{CJKfntef}

\makeatletter
\renewcommand{\@thesubfigure}{\hskip\subfiglabelskip}
\makeatother \allowdisplaybreaks[3]

\begin{document}
\title{QCD axial anomaly enhances the $\eta \eta^\prime$ decay of the hybrid candidate $\eta_1(1855)$}
%

\author{Hua-Xing Chen$^1$}
\email{hxchen@seu.edu.cn}
\author{Niu Su$^1$}
\email{suniu@seu.edu.cn}
\author{Shi-Lin Zhu$^2$}
\email{zhusl@pku.edu.cn}

\affiliation{
$^1$School of Physics, Southeast University, Nanjing 210094, China\\
$^2$School of Physics and Center of High Energy Physics, Peking University, Beijing 100871, China}

\begin{abstract}
We study the hybrid mesons with the exotic quantum number $I^GJ^{PC} = 0^+1^{-+}$ and investigate their decays into the $\eta \eta^\prime$, $a_1(1260) \pi$, $f_1(1285) \eta$, $f_1(1420) \eta$, $K^*(892) \bar K$, $K_1(1270) \bar K$, and $K_1(1400) \bar K$ channels. We find that the QCD axial anomaly enhances the decay width of the $\eta \eta^\prime$ channel although this mode is strongly suppressed by the small $P$-wave phase space. Our results support the interpretation of the $\eta_1(1855)$ recently observed by BESIII as the $\bar s s g$ hybrid meson of $I^GJ^{PC}=0^+1^{-+}$. The QCD axial anomaly ensures the $\eta \eta^\prime$ decay mode to be a characteristic signal of the hybrid nature of the $\eta_1(1855)$.
\end{abstract}
\maketitle
\pagenumbering{arabic}

%
%
%
\section{Introduction}
\label{sec:intro}
%

Very recently, the BESIII collaboration performed a partial wave analysis of the $J/\psi \to \gamma \eta \eta^\prime$ decay, and reported the first observation of an abnormal state with the exotic quantum number $I^GJ^{PC} = 0^+1^{-+}$~\cite{Ablikim:2022zze,Ablikim:2022glj}. It was observed in the $\eta \eta^\prime$ invariant mass spectrum with a statistical significance larger than $19\sigma$. Its mass and width were measured to be
\begin{eqnarray}
\eta_1(1855) &:& M = 1855 \pm 9 ^{+6}_{-1} {\rm~MeV}/c^2 \, ,
\\ \nonumber && \Gamma = 188 \pm 18 ^{+3}_{-8} {\rm~MeV} \, .
\end{eqnarray}

The $\eta_1(1855)$ is the isoscalar partner of the isovector state $\pi_1(1600)$~\cite{E852:1998mbq} or $\pi_1(1400)$~\cite{IHEP-Brussels-LosAlamos-AnnecyLAPP:1988iqi} with $I^GJ^{PC} = 1^-1^{-+}$. These states are of particular interests, since their exotic quantum numbers can not be accessed by conventional $\bar q q$ mesons and may arise from the gluon degree of freedom~\cite{pdg}. One of their possible explanations is the hybrid meson composed of one valence quark and one valence antiquark together with one valence gluon, whose experimental confirmation is a direct test of QCD in the low energy sector.

In the past half century there have been a lot of theoretical studies on hybrid mesons, through the MIT bag model~\cite{Barnes:1977hg,Hasenfratz:1980jv,Chanowitz:1982qj}, flux-tube model~\cite{Isgur:1983wj,Close:1994hc,Page:1998gz}, constituent gluon model~\cite{Horn:1977rq,Szczepaniak:2001rg,Guo:2007sm}, AdS/QCD model~\cite{Andreev:2012hw,Bellantuono:2014lra}, lattice QCD~\cite{Michael:1985ne,Juge:2002br,Lacock:1996ny,MILC:1997usn,Dudek:2009qf,Dudek:2013yja}, and QCD sum
rules~\cite{Balitsky:1982ps,Govaerts:1983ka,Kisslinger:1995yw,Jin:2002rw,Narison:2009vj,Li:2021fwk}, etc. However, their nature still remains elusive due to our poor understanding of the gluon degree of freedom. It is not easy to experimentally identify hybrid mesons unambiguously, and there is currently no definite experimental evidence on their existence. It is also not easy to theoretically define the gluon degree of freedom. A precise definition of the constituent gluon is still lacking, although there have been some proposals to construct glueballs and hybrid mesons using constituent gluons~\cite{Horn:1977rq,Coyne:1980zd,Chanowitz:1980gu,Barnes:1981ac,Cornwall:1982zn,Cho:2015rsa}. We refer to the
reviews~\cite{pdg,Klempt:2007cp,Amsler:2004ps,Bugg:2004xu,Meyer:2010ku,Meyer:2015eta,Chen:2016qju,Briceno:2017max,Ketzer:2019wmd,Jin:2021vct,Chen:2022asf} as well as the recent experimental analyses performed by the Compass and JPAC collaborations~\cite{COMPASS:2018uzl,JPAC:2018zyd} for detailed discussions.

The $\eta_1(1855)$ was interpreted as a hybrid meson in a recent study~\cite{Qiu:2022ktc}. There also exist some other possible explanations, {\it e.g.}, the $\eta_1(1855)$ may be explained as the $K \bar K_1(1400)/K \bar K_1(1270)$ hadronic molecules in Refs.~\cite{Dong:2022cuw,Zhang:2019ykd}. In 2008 we systematically studied the $qs\bar q \bar s$ ($q=u/d$) tetraquark state of $I^GJ^{PC} = 0^+1^{-+}$ using the QCD sum rule method, and we wrote in the abstract~\cite{Chen:2008ne}:`` $\cdots$ these currents lead to mass estimates around $1.8$-$2.1$~GeV, where the uncertainty is due to the mixing of two single currents. Its possible decay modes are $S$-wave $b_1(1235) \eta$ and $b_1(1235) \eta^\prime$, and $P$-wave $KK$, $\eta \eta$, $\eta \eta^\prime$ and $\eta^\prime \eta^\prime$, etc. The decay width is around 150 MeV through a rough estimation.'' Accordingly, the $\eta_1(1855)$ may also be interpreted as a $qs\bar q \bar s$ tetraquark state. Note that we made two mistakes there on its possible decay modes: a) the $b_1(1235) \eta$ and $b_1(1235) \eta^\prime$ decay modes obtained by the Fierz rearrangement should be replaced by the $a_1(1260)\pi$ and $f_1(1285)\eta$ decay modes; b) the $\eta \eta$ and $\eta^\prime \eta^\prime$ decay modes are forbidden due to the Bose-Einstein statistics.

\begin{figure*}[hbtp]
\begin{center}
\subfigure[(a)]{
\includegraphics[width=0.3\textwidth]{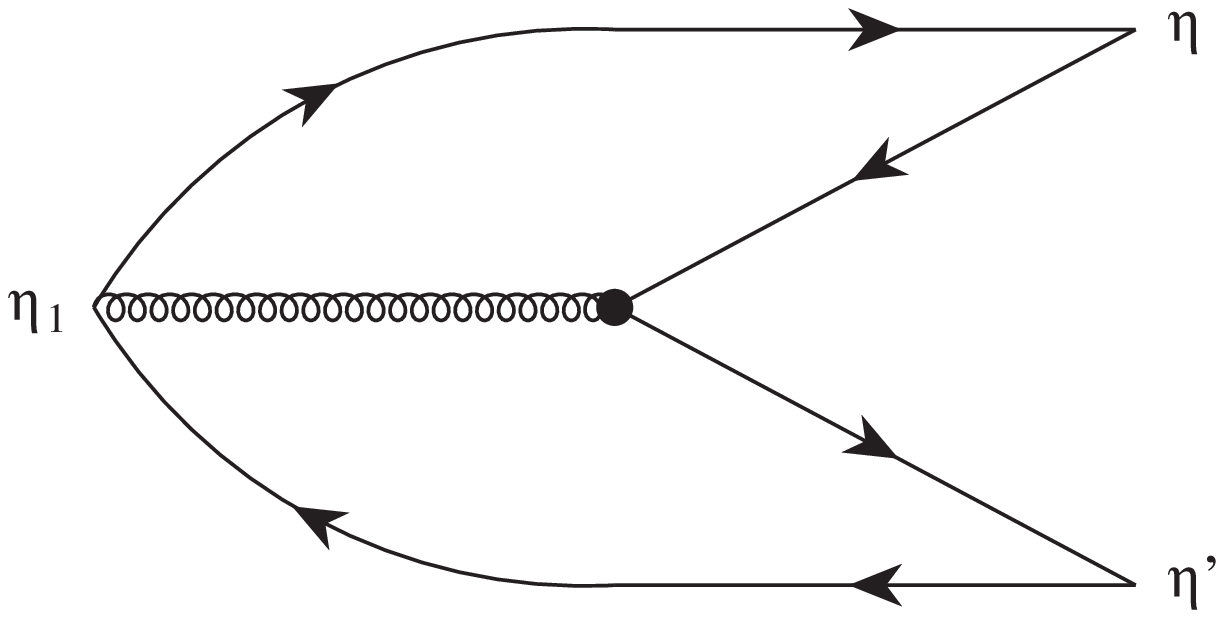}}
~~ \subfigure[(b)]{
\includegraphics[width=0.3\textwidth]{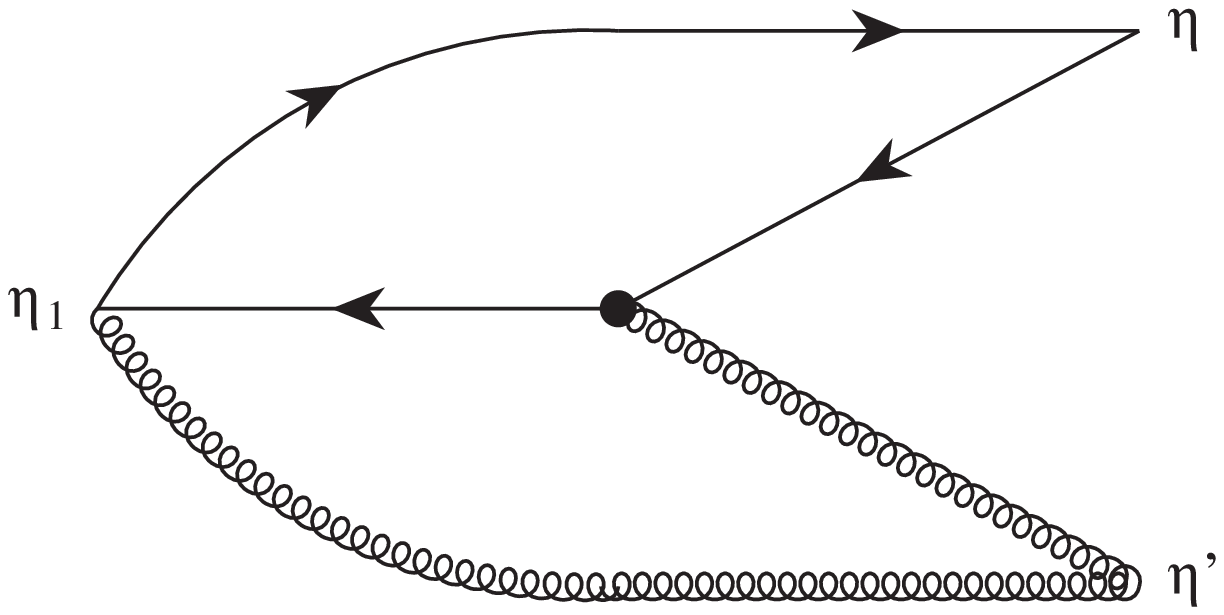}}
~~ \subfigure[(c)]{
\includegraphics[width=0.3\textwidth]{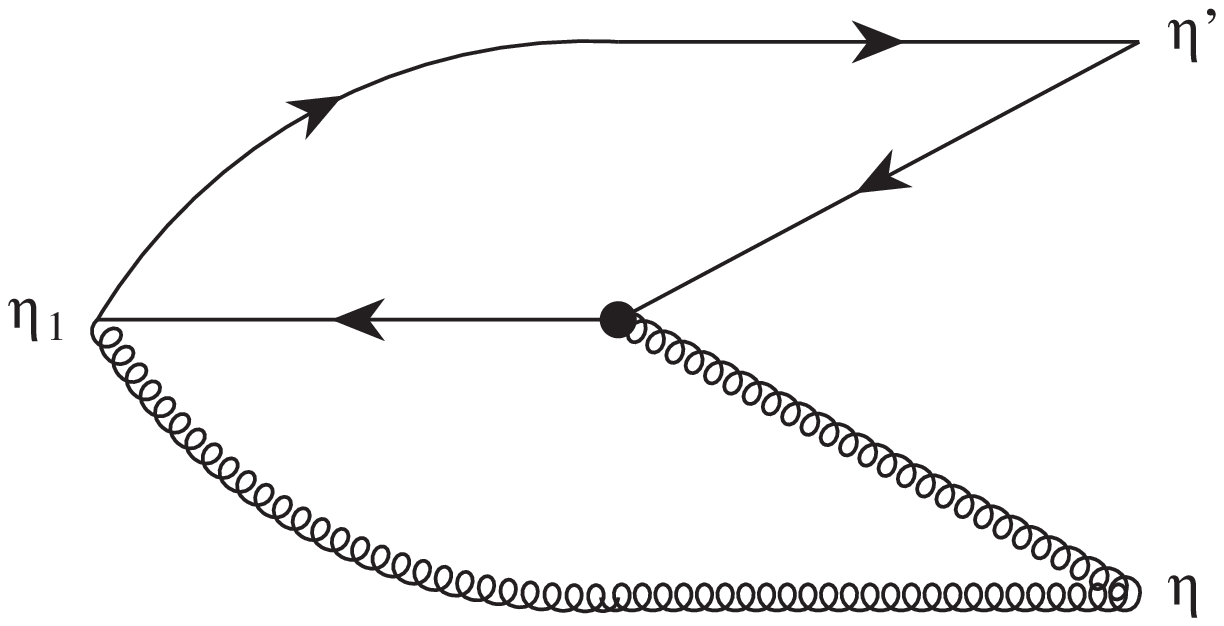}}
\end{center}
\caption{Possible decay mechanisms of the hybrid meson through (a) the normal process with one quark-antiquark pair excited from the valence gluon as well as the abnormal processes with the (b) $\eta^\prime$ and (c) $\eta$ mesons produced by the QCD axial anomaly.}
\label{fig:decay}
\end{figure*}

In 2010 we systematically studied decay properties of the $\bar q q g$ ($q=u/d$) hybrid meson with $I^GJ^{PC} = 0^+1^{-+}$ using the methods of QCD sum rules and light-cone sum rules, where we pointed out its $\eta \eta^\prime$ decay mode~\cite{Huang:2010dc,Chen:2010ic}. This is just the discovery channel of the $\eta_1(1855)$ observed by BESIII~\cite{Ablikim:2022zze,Ablikim:2022glj}. However, the partial width of this decay mode was calculated to be quite small in Ref.~\cite{Chen:2010ic}, where we only took into account the normal decay process depicted in Fig.~\ref{fig:decay}(a) with one quark-antiquark pair excited from the valence gluon.

There exists a decades-long puzzle in the search of the hybrid mesons. Several experimental groups reported the isovector states $\pi_1(1400)$~\cite{IHEP-Brussels-LosAlamos-AnnecyLAPP:1988iqi} and $\pi_1(1600)$~\cite{E852:1998mbq} in the $\eta \pi$ and $\eta^\prime \pi$ channels, respectively. However, a ``selection rule'' was proposed in Refs.~\cite{Page:1998gz,Page:1996rj} through the popular flux tube model that the $J^{PC} = 1^{-+}$ hybrid meson does not decay into two $S$-wave ground-state mesons. In other words, the $\eta \pi$ and $\eta^\prime \pi$ modes are strictly forbidden in the flux tube model, which is in strong contrast of the experimental fact that these two modes together with the $\eta\eta^\prime$ are the discovery modes of $\pi_1(1400)$, $\pi_1(1600)$, and $\eta_1(1885)$, respectively. The above selection rule can be interpreted from the Feynman diagram depicted in Fig.~\ref{fig:decay}(a) to some extent, where the leading-order perturbative contribution vanishes in the chiral limit. In our previous QCD sum rule calculation, the small $\eta \pi$ decay width arises from two chirality-violating sources: the $up/down$ current quark mass and the non-perturbative QCD condensates.

In the present work, we shall further investigate the anomaly-assisted (or ``abnormal'') decay processes depicted in Fig.~\ref{fig:decay}(b,c) with the $\eta^{(\prime)}$ mesons produced by the QCD axial anomaly. We shall find that these diagrams enhance the $\eta \eta^\prime$ decay mode. We want to emphasize that this is the characteristic decay mode of the hybrid mesons if we consider the QCD axial anomaly~\footnote{A similar mechanism was investigated in [Phys. Lett. B \textbf{214}, 463-466 (1988)] to study the decays of the $J^{PC}=1^{-+}$ hybrid mesons. We thank Professor Jean-Marie Frere for informing us this work after our manuscript was posted on the arXiv.}.

In this paper we shall also update our previous calculations of Ref.~\cite{Chen:2010ic} to investigate the $a_1(1260) \pi$, $f_1(1285) \eta$, $f_1(1420) \eta$, $K^*(892) \bar K$, $K_1(1270) \bar K$, and $K_1(1400) \bar K$ decay channels for both the $\bar q q g$ and $\bar s s g$ hybrid mesons of $I^GJ^{PC} = 0^+1^{-+}$. We shall find that the $\eta_1(1855)$ recently observed by BESIII can be well explained as the $\bar s s g$ hybrid meson. Moreover, the connection with the QCD axial anomaly strongly suggests the hybrid nature of the $\eta_1(1855)$.

This paper is organized as follows. In Sec.~\ref{sec:etaetap}, we investigate the QCD axial anomaly and study its contribution to the $\eta_1(1855) \to \eta \eta^\prime$ decay channel, based on the interpretation of the $\eta_1(1855)$ as the $\bar s s g$ hybrid meson with $I^GJ^{PC} = 0^+1^{-+}$. In Sec.~\ref{sec:others} we study the $\eta_1(1855)$ decay into the $K^*(892) \bar K$ and $K_1(1270) \bar K$ channels. The $a_1(1260) \pi$, $f_1(1285) \eta$, $f_1(1420) \eta$, and $K_1(1400) \bar K$ decay channels are investigated in Sec.~\ref{sec:summary} for both the $\bar s s g$ and $\bar q q g$ hybrid mesons together with a summary of the present study.

\section{The $\eta \eta^\prime$ decay channel}
\label{sec:etaetap}

In this section we assume the $\eta_1(1855)$ to be the $\bar s s g$ hybrid meson of $I^GJ^{PC} = 0^+1^{-+}$, and study its decay into the $\eta \eta^\prime$ channel. We shall concentrate on the abnormal decay process depicted in Fig.~\ref{fig:decay}(b) with the $\eta^\prime$ meson produced by the QCD axial anomaly. The other two decay processes depicted in Fig.~\ref{fig:decay}(a,c) will be shortly discussed at the end of this section.

\subsection{QCD axial anomaly}

The conversion of the gluons into the $\eta$ and $\eta^\prime$ mesons is described by an amplitude relating to the $SU(3)$ symmetry and the QCD $U(1)_A$ anomaly. Its matrix elements can be written as~\cite{Voloshin:1980zf,Akhoury:1987ed,Castoldi:1988dm,Chao:1989yp,Ball:1995zv,Ali:1997ex}:
\begin{eqnarray}
\langle 0| {\alpha_s \over 4 \pi} G^{\alpha\beta}_n \tilde G_{n,\alpha\beta} | \eta \rangle &=& m^2_{\eta} f_{\eta} \, ,
\label{eq:anomaly1}
\\
\langle 0| {\alpha_s \over 4 \pi} G^{\alpha\beta}_n \tilde G_{n,\alpha\beta} | \eta^\prime \rangle &=& m^2_{\eta^\prime} f_{\eta^\prime} \, ,
\label{eq:anomaly2}
\end{eqnarray}
where $G^{\alpha\beta}_n$ is the gluon field strength tensor and $\tilde G_{n,\alpha\beta} = G_n^{\mu\nu} \times \epsilon_{\mu\nu\alpha\beta}/2$ is its dual field, with $n=1\cdots8$ the color index and $\alpha,\beta$ the Lorentz indices. The two decay constants $f_{\eta}$ and $f_{\eta^\prime}$ are defined as
\begin{eqnarray}
f_{\eta} &=& {f_8 \over \sqrt6} \cos\theta_8 - {f_0 \over \sqrt3} \sin\theta_0 \, ,
\\ \nonumber
f_{\eta^\prime} &=& {f_8 \over \sqrt6} \sin\theta_8 + {f_0 \over \sqrt3} \cos\theta_0 \, .
\end{eqnarray}

In the above expressions we have followed Refs.~\cite{Leutwyler:1997yr,Kaiser:1998ds,Escribano:2005qq,Escribano:2015nra,Escribano:2015yup,Schechter:1992iz,Kiselev:1992ms,Herrera-Siklody:1997pgy,Bass:2018xmz,Bali:2021qem} and used the two-angle mixing formalism to describe the $\eta$ and $\eta^\prime$ mesons:
\begin{eqnarray}
|\eta\rangle &=& \cos\theta_8 |\eta_8\rangle - \sin \theta_0 | \eta_0 \rangle + \cdots \, ,
\\ \nonumber
|\eta^\prime\rangle &=& \sin\theta_8 |\eta_8\rangle + \cos \theta_0 | \eta_0 \rangle + \cdots \, ,
\end{eqnarray}
where
\begin{eqnarray}
|\eta_8 \rangle &=& | u \bar u + d \bar d - 2 s \bar s \rangle/\sqrt6 \, ,
\\ \nonumber
|\eta_0 \rangle &=& | u \bar u + d \bar d + s \bar s \rangle/\sqrt3 \, ,
\end{eqnarray}
and $\cdots$ are contributed by some other components such as the pseudoscalar glueball and charmonium, etc.

The octet and singlet axial-vector currents mesons are defined as
\begin{eqnarray}
A_\mu^{8} &=& \left( { \bar u \gamma_\mu \gamma_5 u +  \bar d \gamma_\mu \gamma_5 d - 2 \bar s \gamma_\mu \gamma_5 s }\right)/\sqrt{12} \, ,
\\ \nonumber
A_\mu^{0} &=& \left( { \bar u \gamma_\mu \gamma_5 u +  \bar d \gamma_\mu \gamma_5 d + \bar s \gamma_\mu \gamma_5 s }\right)/\sqrt6 \, ,
\end{eqnarray}
which couple to the $\eta$ and $\eta^\prime$ mesons through
\begin{equation}
\langle0| A_\mu^a | P(k) \rangle = i k_\mu f_P^a \, ,
\end{equation}
with the matrix for the decay constants $f_P^a$ ($a=8,0;\,P=\eta,\eta^\prime$) defined as
\begin{equation}
\left(\begin{array}{cc}
f_\eta^8 & f_\eta^0
\\
f_{\eta^\prime}^8 & f_{\eta^\prime}^0
\end{array}\right)
=
\left(\begin{array}{cc}
f_8 \cos\theta_8 & - f_0 \sin\theta_0
\\
f_8 \sin\theta_8 &   f_0 \cos\theta_0
\end{array}\right) \, .
\end{equation}

To simply our calculations, we further construct the interpolating currents
\begin{eqnarray}
J_\mu^\eta = A_\mu^8 + t_\eta A_\mu^0 \, ,
\\ \nonumber
J_\mu^{\eta\prime} = A_\mu^8 +  t_{\eta^\prime} A_\mu^0 \, ,
\end{eqnarray}
which separately couple to the $\eta$ and $\eta^\prime$ mesons through
\begin{eqnarray}
\nonumber \langle0| J_\mu^\eta |\eta(k) \rangle &=& i k_\mu g_\eta \, ,
\\
\langle0| J_\mu^{\eta\prime} |\eta^\prime(k) \rangle &=& i k_\mu g_{\eta^\prime} \, ,
\\ \nonumber
\langle0| J_\mu^\eta |\eta^\prime(k) \rangle &=& \langle0| J_\mu^{\eta\prime} |\eta(k) \rangle = 0 \, ,
\end{eqnarray}
with the parameters
\begin{eqnarray}
\nonumber g_\eta &=& f_\eta^8 - f_\eta^0 f_{\eta^\prime}^8/f_{\eta^\prime}^0 \, ,
\\ g_{\eta^\prime} &=& f_{\eta^\prime}^8 - f_{\eta^\prime}^0 f_{\eta}^8/f_{\eta}^0 \, ,
\\ \nonumber t_\eta &=& - f_{\eta^\prime}^8/f_{\eta^\prime}^0 \, ,
\\ \nonumber t_{\eta^\prime} &=& - f_{\eta}^8/f_{\eta}^0 \, .
\end{eqnarray}
The following values will be used for the parameters contained in the above equations~\cite{Ali:1997ex,Feldmann:1997vc}:
\begin{eqnarray}
\nonumber \theta_8 &=& - 22.2^\circ \, ,
\\ \nonumber \theta_0 &=& - 9.1^\circ \, ,
\\ f_8 &=& 168 {\rm~MeV} \, ,
\label{parameter1}
\\ \nonumber f_0 &=& 157 {\rm~MeV} \, .
\end{eqnarray}

At the same time we can reorganize the currents $A_\mu^{8}$, $A_\mu^{0}$, $J_\mu^\eta$, and $J_\mu^{\eta\prime}$ to be
\begin{eqnarray}
J_\mu^{\bar q q} &=& \left( { \bar u \gamma_\mu \gamma_5 u +  \bar d \gamma_\mu \gamma_5 d }\right)/\sqrt2 \, ,
\\ \nonumber
J_\mu^{\bar s s} &=& \bar s \gamma_\mu \gamma_5 s \, .
\end{eqnarray}
They can also couple to the $f_1(1285)$ and $f_1(1420)$ through
\begin{eqnarray}
\langle0| J_\mu^{\bar q q} |f_1(1285)\rangle &=& f_{f_1(1285)} m_{f_1(1285)} \epsilon_\mu \, ,
\\ \nonumber \langle0| J_\mu^{\bar s s} |f_1(1420)\rangle &=& f_{f_1(1420)} m_{f_1(1420)} \epsilon_\mu \, ,
\end{eqnarray}
where the $f_1(1285)$ and $f_1(1420)$ mesons are assumed to be predominated by their $\bar q q$ and $\bar s s$ components~\cite{pdg,Yang:2007zt}. The decay constants $f_{f_1(1285)}$ and $f_{f_1(1420)}$ were evaluated in Ref.~\cite{Yang:2007zt} to be:
\begin{eqnarray}
f_{f_1(1285)} &=& 173 \pm 23{\rm~MeV} \, ,
\\ \nonumber f_{f_1(1420)} &=& 217 \pm 27{\rm~MeV} \, .
\end{eqnarray}

\subsection{Three-point correlation function}

The QCD sum rule method has proven to be a powerful and successful non-perturbative method for the past decades~\cite{Shifman:1978bx,Reinders:1984sr}, and we applied it to study three-gluon glueballs and double-gluon hybrid mesons recently~\cite{Chen:2021cjr,Chen:2021bck,Chen:2021smz}. We consider the following three-point correlation function to study the decay process depicted in Fig.~\ref{fig:decay}(b):
\begin{equation}
T^b_{\mu\nu}(p, k, q) = \int d^4x e^{-ikx} \langle 0 | {\mathbb T} J_\mu^{\eta_1}(0) J_\nu^{\eta\dagger}(x) | \eta^\prime(q) \rangle \, ,
\label{eq:correlation}
\end{equation}
where $p$, $k$, and $q$ are the momenta of the $\eta_1(1855)$, $\eta$, and $\eta^\prime$, respectively. We also need Eq.~(\ref{eq:anomaly2}) to relate $\eta^\prime$ to the QCD axial anomaly.

The $\bar s s g$ hybrid current $J_\mu^{\eta_1}$ of $I^GJ^{PC}=0^+1^{-+}$ is defined as:
\begin{equation}
J^{\eta_1}_{\mu} = \bar s_a \gamma^\nu s_b { \lambda^{ab}_n \over 2 } g_s G^n_{\mu\nu} \, ,
\end{equation}
which couples to the $\eta_1(1855)$ through
\begin{equation}
\langle 0 | J^{\eta_1}_{\mu} | \eta_1(\epsilon,p) \rangle = \sqrt 2 f_{\eta_1} m_{\eta_1}^3 \epsilon_\mu \, .
\end{equation}
We choose the value of $f_{\eta_1}$ to be the same as that of the $\bar q q g$ hybrid meson~\cite{Narison:1989aq,Zhu:1999wg,Zhu:1998bm}:
\begin{equation}
f_{\eta_1} \approx 0.026 {\rm~GeV} \, , \label{parameter3}
\end{equation}
with the uncertainty $\pm15\%$.

At the hadronic level we express Eq.~(\ref{eq:correlation}) as
\begin{eqnarray}
&& T^b_{\mu\nu}(p, k, q)
\label{eq:hadron}
\\ \nonumber &=& g^{b}_{\eta\eta^\prime} k_\mu k_\nu ~ { \sqrt 2 f_{\eta_1} m_{\eta_1}^3 g_{\eta} \over (m_{\eta_1}^2 - p^2) (m_{\eta}^2 - k^2)}
\\ \nonumber &+& g_{f_1\eta} \theta_s \left( g_{\mu\nu} - {k_\mu k_\nu \over m_{f_1}^2} \right) { \sqrt 2 f_{\eta_1} m_{\eta_1}^3 f_{f_1} m_{f_1} \over (m_{\eta_1}^2 - p^2) (m_{f_1}^2 - k^2)} + \cdots ,
\end{eqnarray}
where $\theta_s = -{1 / \sqrt3} + { t_\eta / \sqrt6}$ describes the $s \bar s$ component contained in the current $J_\nu^{\eta}$. The coupling constants $g^b_{\eta\eta^\prime}$ and $g_{f_1\eta}$ are defined through the Lagrangians
\begin{eqnarray}
\mathcal{L}^b_{\eta\eta^\prime} &=& i g^b_{\eta\eta^\prime} \eta_1^\mu (\partial_\mu \eta) \eta^\prime \, ,
\\ \nonumber \mathcal{L}_{f_1\eta} &=& g_{f_1\eta} \eta_1^\mu f_{1\mu} \eta \, .
\end{eqnarray}

At the quark-gluon level we calculate Eq.~(\ref{eq:correlation}) using the method of operator product expansion (OPE). Based on Eq.~(\ref{eq:anomaly2}), we calculate the decay process depicted in Fig.~\ref{fig:decay}(b) to be
\begin{eqnarray}
&& T^b_{\mu\nu}(p, k, q)
\label{eq:OPE}
\\ \nonumber &=&
\theta_s k_\mu k_\nu ~ \Big( - {2 m^2_{\eta^\prime} f_{\eta^\prime} \over 3 k^2 } - {4\pi^2 m^2_{\eta^\prime} f_{\eta^\prime} m_s \langle \bar s s\rangle \over 3k^6} \Big)
\\ \nonumber &+&
\theta_s \left( g_{\mu\nu} - {k_\mu k_\nu \over k^2} \right) {4\pi^2 m^2_{\eta^\prime} f_{\eta^\prime} m_s \langle \bar s s\rangle \over 9 k^4 } + \cdots .
\end{eqnarray}

Paying attention to the Lorentz structure $k_\mu k_\nu$, we perform the Borel transformation to Eq.~(\ref{eq:hadron}) and Eq.~(\ref{eq:OPE}) at the limit of $p^2 \sim k^2 \rightarrow \infty$, and obtain:
\begin{eqnarray}
&& \nonumber g^b_{\eta\eta^\prime} {\sqrt 2 m_{\eta_1}^3 f_{\eta_1} g_\eta \over m_{\eta_1}^2 - m_\eta^2} \left( e^{-m_{\eta}^2/M_B^2} -  e^{-m_{\eta_1}^2/M_B^2} \right)
\\ &=& {2 \theta_s m^2_{\eta^\prime} f_{\eta^\prime} \over 3 } + {2 \pi^2 \theta_s m^2_{\eta^\prime} f_{\eta^\prime} m_s \langle \bar s s\rangle \over 3} {1 \over M_B^4} \, .
\label{eq:etaetap}
\end{eqnarray}
There are only two terms left, and the latter one proportional to $m_s$ is negligible compared to the former one.

\subsection{Numerical analysis}

In the present study we use the following values for various quark and gluon parameters when performing numerical analyses~\cite{pdg,Ovchinnikov:1988gk,Yang:1993bp,Ellis:1996xc,Jamin:2002ev,Ioffe:2002be,Gimenez:2005nt,Narison:2011xe,Narison:2018dcr}:
%
\begin{eqnarray}
\nonumber m_s(1\mbox{ GeV}) &=& \left(93 ^{+11}_{-~5}\right) \times 1.35 \mbox{ MeV} \, ,
\\ \nonumber  \langle\bar qq \rangle &=& -(0.24 \pm 0.01)^3 \mbox{ GeV}^3 \, ,
\\ \nonumber  \langle\bar ss \rangle &=& (0.8\pm 0.1)\times \langle\bar qq \rangle \, ,
\\ \langle g_s\bar q\sigma G q\rangle &=& - M_0^2\times\langle\bar qq\rangle \, ,
\label{condensates}
\\ \nonumber \langle g_s\bar s\sigma G s\rangle &=& - M_0^2\times\langle\bar ss\rangle \, ,
\\ \nonumber M_0^2 &=& (0.8 \pm 0.2) \mbox{ GeV}^2 \, ,
\\ \nonumber \langle \alpha_s GG\rangle &=& (6.35 \pm 0.35) \times 10^{-2} \mbox{ GeV}^4 \, ,
\\ \nonumber \langle g_s^3G^3\rangle &=& \langle \alpha_s GG\rangle \times (8.2 \pm 1.0) \mbox{ GeV}^2 \, .
\end{eqnarray}
%

\begin{figure}[hbt]
\begin{center}
\scalebox{0.6}{\includegraphics{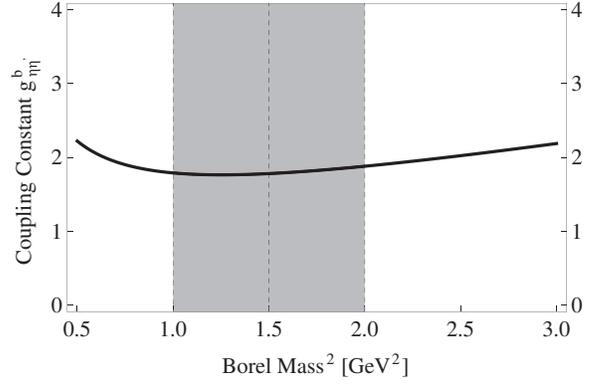}} \caption{The coupling constant $g^b_{\eta\eta^\prime}$ as a function of the Borel mass $M_B^2$, calculated using Eq.~({\ref{eq:etaetap})}.}
\label{fig:etaetap}
\end{center}
\end{figure}

Together with Eqs.~(\ref{parameter1}) and (\ref{parameter3}), we calculate the coupling constant $g^b_{\eta\eta^\prime}$ through Eq.~({\ref{eq:etaetap})}. We show it in Fig.~\ref{fig:etaetap} as a function of the Borel mass $M_B$. We obtain inside the Borel window $M_B^2 = 1.5 \pm 0.5$~GeV$^2$ that
\begin{equation}
| g^b_{\eta \eta^\prime} | = 1.78 \, .
\end{equation}
The above coupling constant is non-negligible. We further derive
\begin{equation}
\Gamma(\eta_1(1855) \xrightarrow{b} \eta \eta^\prime) = 1.80^{+0.72}_{-0.44}{\rm~MeV} \, .
\label{eq:etaetapwidth}
\end{equation}
Its uncertainty is due to the Borel mass and various parameters given in Eqs.~(\ref{parameter1}), (\ref{parameter3}), and (\ref{condensates}). Because some of the parameters do not have uncertainties, the uncertainties of our results should be eventually larger, {\it i.e.}, to be two times larger or smaller.

\subsection{Other processes}

Similarly, we investigate the abnormal decay process depicted in Fig.~\ref{fig:decay}(c) through the three-point correlation function
\begin{equation}
T^c_{\mu\nu}(p, k, q) = \int d^4x e^{-ikx} \langle 0 | {\mathbb T} J_\mu^{\eta_1}(0) J_\nu^{\eta^\prime\dagger}(x) | \eta(q) \rangle \, .
\end{equation}
In this case the $\eta$ meson is produced by the QCD axial anomaly through Eq.~(\ref{eq:anomaly1}), and the coupling constant $g^c_{\eta\eta^\prime}$ is defined through the Lagrangian
\begin{equation}
\mathcal{L}^c_{\eta\eta^\prime} = i g^c_{\eta\eta^\prime} \eta_1^\mu (\partial_\mu \eta^\prime) \eta \, .
\end{equation}
We follow the same procedures to obtain
\begin{equation}
|g^c_{\eta \eta^\prime}| = 0.53 \, ,
\end{equation}
which is used to further derive
\begin{equation}
\Gamma(\eta_1(1855) \xrightarrow{c} \eta \eta^\prime) = 0.74^{+0.42}_{-0.18}{\rm~MeV} \, .
\label{eq:etaetapwidth2}
\end{equation}
This value is at the same level as that given in Eq.~(\ref{eq:etaetapwidth}). However, we are not able to exactly derive the phase angle between their amplitudes, so we shall simply sum over these partial decay widths in Sec.~\ref{sec:summary}.

The normal decay process depicted in Fig.~\ref{fig:decay}(a), with one quark-antiquark pair excited from the valence gluon, has been studied in Ref.~\cite{Chen:2010ic} for the $\bar q q g$ hybrid meson. We replace the light $up/down$ quarks by the $strange$ quarks, and evaluate its partial decay width to be
\begin{equation}
\Gamma(\eta_1(1855) \xrightarrow{a} \eta \eta^\prime) = 0.68^{+2.26}_{-0.36}{\rm~MeV} \, .
\end{equation}
This value is smaller than that given in Eq.~(\ref{eq:etaetapwidth}), so the abnormal decay process depicted in Fig.~\ref{fig:decay}(b) with the $\eta^{\prime}$ meson produced by the QCD axial anomaly does enhance the $\eta \eta^\prime$ decay mode. Note that this value 0.68~MeV is larger than that obtained in Ref.~\cite{Chen:2010ic}, because the gluon condensates $\langle \alpha_s GG\rangle$ and $\langle g_s^3G^3\rangle$ used here are larger than those used there.

Besides the Lorentz structure $k_\mu k_\nu$ investigated above, the three-point correlation functions $T^b_{\mu\nu}(p, k, q)$ given in Eq.~(\ref{eq:hadron}) and Eq.~(\ref{eq:OPE}) also contain the Lorentz structure $\left( g_{\mu\nu} - {k_\mu k_\nu / k^2} \right)$. These components can be used to study the contribution of the QCD axial anomaly to the $f_1 \eta$ decay channel. Assuming the mass of $\eta_1$ to be around 2100~MeV, its decay into the $f_1(1420) \eta$ channel is kinematically allowed. However, the partial decay width of this mode is evaluated to be quite small, so we shall not discuss it any more.

\section{Other decay channels}
\label{sec:others}

\subsection{The $K^* \bar K$ decay channel}
\label{sec:KsK}

In this subsection we assume the $\eta_1(1855)$ to be the $\bar s s g$ hybrid meson of $I^GJ^{PC} = 0^+1^{-+}$, and study its decay into the $K^*(892) \bar K$ channel through the normal decay process depicted in Fig.~\ref{fig:decay}(a). We follow the same procedures as those used in Ref.~\cite{Chen:2010ic} for the $\pi_1 \to \rho \pi$ decay, and study the three-point correlation function
\begin{eqnarray}
&& T^{K^* \bar K}_{\mu\nu}(p, k, q)
\label{eq:K1K}
\\ \nonumber &=& \int d^4x d^4y e^{i k x} e ^{i q y} \langle0|{\mathbb T} J^{K^*}_{\nu}(x) J^{\bar K}(y) J_\mu^{\eta_1\dagger}(0) |0\rangle \, .
\end{eqnarray}
The currents $J^{K^*}_{\nu}$ and $J^{\bar K}$ are defined as
\begin{eqnarray}
J^{K^*}_{\nu} &=& \bar q \gamma_\nu s \, ,
\\ \nonumber J^{\bar K} &=& \bar s \gamma_5 q \, ,
\end{eqnarray}
which couple to the $K^*(892)$ and $\bar K$ mesons through
\begin{eqnarray}
\langle0| J^{K^*}_{\nu} | K^*(\epsilon, k)\rangle &=& f_{K^*} m_{K^*} \epsilon_\nu \, ,
\\ \nonumber \langle0| J^{\bar K} | \bar K(q)\rangle &=& \lambda_K \, .
\label{eq:K}
\end{eqnarray}

We use the following values for the parameters in the above equations~\cite{pdg,Becirevic:2003pn}:
\begin{eqnarray}
\nonumber f_{K^*} &=& 217 {\rm~MeV} \, ,
\\ \lambda_K &=& {i(\langle\bar q q\rangle + \langle \bar s s \rangle) / f_K} \, ,
\\ \nonumber f_{K^+} &=& 155.7 {\rm~MeV} \, ,
\end{eqnarray}
and obtain
\begin{equation}
\Gamma(\eta_1(1855) \to K^*(892) \bar K + c.c.) = 98.1^{+82.9}_{-60.4} {\rm~MeV} \, .
\end{equation}

\subsection{The $K_1(1270) \bar K$ decay channel}

In this subsection we assume the $\eta_1(1855)$ to be the $\bar s s g$ hybrid meson of $I^GJ^{PC} = 0^+1^{-+}$, and study its decay into the $K_1(1270) \bar K$ channel through the normal decay process depicted in Fig.~\ref{fig:decay}(a). We follow the same procedures as those used in Ref.~\cite{Chen:2010ic} for the $\pi_1 \to f_1 \pi$ decay, and study the three-point correlation function
\begin{eqnarray}
&& T^{K_1 \bar K}_{\mu\nu}(p, k, q)
\label{eq:K1K}
\\ \nonumber &=& \int d^4x d^4y e^{i k x} e ^{i q y} \langle0|{\mathbb T} J^{K_1}_{\nu}(x) J^{\bar K}(y) J_\mu^{\eta_1\dagger}(0) |0\rangle \, .
\end{eqnarray}
The current $J^{K_1}_{\nu}$ is defined as
\begin{equation}
J^{K_1}_{\nu} = \bar q \gamma_\nu \gamma_5 s \, ,
\end{equation}
which couples to both the $K_1 \equiv K_1(1270)$ and $K_1^\prime \equiv K_1(1400)$ mesons through
\begin{eqnarray}
\langle0| J^{K_1}_{\nu} | K_1(\epsilon, k)\rangle &=& f_{K_1} m_{K_1} \epsilon_\nu \, ,
\\ \nonumber \langle0| J^{K_1}_{\nu} | K_1^\prime(\epsilon, k)\rangle &=& f_{K_1^\prime} m_{K_1^\prime} \epsilon_\nu \, .
\end{eqnarray}

We use the following values for the parameters in the above equations~\cite{pdg,Yang:2007zt}:
\begin{eqnarray}
f_{K_1} &=& 197 \pm 15 {\rm~MeV} \, ,
\\ \nonumber f_{K_1^\prime} &=& 217 \pm 27 {\rm~MeV} \, ,
\end{eqnarray}
and obtain
\begin{equation}
\Gamma(\eta_1(1855) \to K_1(1270) \bar K + c.c.) = 30.4^{+40.1}_{-19.3} {\rm~MeV} \, .
\end{equation}

%
\section{Summary and Discussions}
\label{sec:summary}
%

\begin{table*}[tbh]
\begin{center}
\renewcommand{\arraystretch}{1.5}
\caption{Partial decay widths of the $\bar q q g$ ($q=u/d$) and $\bar s s g$ hybrid mesons with $I^GJ^{PC}=0^+1^{-+}$. $\Gamma(\eta_1(1855) \xrightarrow{a,b,c} \eta \eta^\prime)$ denote the partial decay widths of the processes depicted in Fig.~\ref{fig:decay}(a,b,c), respectively. Masses and widths are in units of MeV.}
\begin{tabular}{c|c|c|c|c}
\hline \hline
\multirow{2}*{Channel}                    & \multicolumn{2}{c|}{Scheme-I}                                     & \multicolumn{2}{c}{Scheme-II}
\\ \cline{2-5}                            & ~~$\bar q q g$ with $M\approx1600$~~ & ~~$\bar s s g$ with $M=1855$~~ & ~~$\bar q q g$ with $M=1855$~~ & ~~$\bar s s g$ with $M\approx2100$~~
\\ \hline \hline
$\eta_1 \xrightarrow{a} \eta \eta^\prime$ & $0.02^{+0.04}_{-0.02}$               & $0.7^{+2.3}_{-0.4}$            & $0.2^{+0.5}_{-0.1}$            & $2.5^{+1.8}_{-2.0}$
\\ \hline
$\eta_1 \xrightarrow{b} \eta \eta^\prime$ & $2.3^{+3.2}_{-2.4}$                  & $1.8^{+0.7}_{-0.4}$            & $9.0^{+3.6}_{-2.2}$            & $2.5^{+1.0}_{-0.7}$
\\ \hline
$\eta_1 \xrightarrow{c} \eta \eta^\prime$ & $0.4^{+0.5}_{-0.4}$                  & $0.7^{+0.4}_{-0.2}$            & $1.6^{+0.9}_{-0.4}$            & $1.1^{+0.7}_{-0.3}$
\\ \hline
$\eta_1 \rightarrow a_1(1260) \pi$        & $50.3^{+62.9}_{-40.6}$               & --                             & $145^{+182}_{-~88}$            & --
\\ \hline
$\eta_1 \rightarrow f_1(1285) \eta$       & --                                   & --                             & $11.8^{+17.1}_{-~8.3}$         & --
\\ \hline
$\eta_1 \rightarrow f_1(1420) \eta$       & --                                   & --                             & --                             & $24.4^{+55.2}_{-23.8}$
\\ \hline
$\eta_1 \rightarrow K^*(892)\bar K+c.c.$  & $25.6^{+25.3}_{-20.7}$               & $98.1^{+82.9}_{-60.4}$         & $63.0^{+51.7}_{-37.0}$         & $154^{+137.3}_{-~97.7}$
\\ \hline
$\eta_1 \rightarrow K_1(1270)\bar K+c.c.$ & --                                   & $30.4^{+40.1}_{-19.3}$         & $20.2^{+26.0}_{-12.2}$         & $76.7^{+120.7}_{-~53.8}$
\\ \hline
$\eta_1 \rightarrow K_1(1400)\bar K+c.c.$ & --                                   & --                             & --                             & $79.1^{+167.1}_{-~63.0}$
\\ \hline \hline
Sum                                       & $79^{+92}_{-64}$                     & $132^{+126}_{-~81}$            & $251^{+282}_{-148}$            & $340^{+484}_{-241}$
\\ \hline \hline
\end{tabular}
\label{tab:result}
\end{center}
\end{table*}

In this paper we apply the method of QCD sum rules to study the $\eta_1(1855)$ recently observed by BESIII as the $\bar s s g$ hybrid meson of $I^GJ^{PC}=0^+1^{-+}$, and calculate its partial decay widths into the $\eta \eta^\prime$, $K^*(892) \bar K$, and $K_1(1270) \bar K$ channels. We have considered the normal decay process depicted in Fig.~\ref{fig:decay}(a) with one quark-antiquark pair excited from the valence gluon. We have also considered the abnormal decay processes depicted in Fig.~\ref{fig:decay}(b,c) with the $\eta^{(\prime)}$ mesons produced by the QCD axial anomaly. The results are
\begin{eqnarray*}
\Gamma(\eta_1(1855) \xrightarrow{a} \eta \eta^\prime) &=& 0.7^{+2.3}_{-0.4}{\rm~MeV} \, ,
\\
\Gamma(\eta_1(1855) \xrightarrow{b} \eta \eta^\prime) &=& 1.8^{+0.7}_{-0.4}{\rm~MeV} \, ,
\\
\Gamma(\eta_1(1855) \xrightarrow{c} \eta \eta^\prime) &=& 0.7^{+0.4}_{-0.2}{\rm~MeV} \, ,
\\
\Gamma(\eta_1(1855) \to K^*(892) \bar K + c.c.) &=& 98.1^{+82.9}_{-60.4}{\rm~MeV} \, ,
\\
\Gamma(\eta_1(1855) \to K_1(1270) \bar K + c.c.) &=& 30.4^{+40.1}_{-19.3} {\rm~MeV} \, ,
\end{eqnarray*}
where $\Gamma(\eta_1(1855) \xrightarrow{a,b,c} \eta \eta^\prime)$ denote the partial decay widths of the processes depicted in Fig.~\ref{fig:decay}(a,b,c), respectively. We simply sum over them to obtain $\Gamma_{\rm sum} \approx 132^{+126}_{-~81}$~MeV. Considering that there still exist some other possible decay channels, our results support the interpretation of the $\eta_1(1855)$ as the $\bar s s g$ hybrid meson of $I^GJ^{PC}=0^+1^{-+}$. Moreover, the decay processes depicted in Fig.~\ref{fig:decay}(b,c) with the $\eta^{(\prime)}$ mesons produced by the QCD axial anomaly enhance the $\eta \eta^\prime$ decay mode, and this possible relationship to the QCD axial anomaly implies the hybrid nature of the $\eta_1(1855)$.

It is interesting to compare our results to those of Ref.~\cite{Dong:2022cuw}. In this paper we explain the $\eta_1(1855)$ as the $\bar s s g$ hybrid meson of $I^GJ^{PC}=0^+1^{-+}$, and its partial decay width into the $K^*(892) \bar K$ channel was estimated to be around 98.1~MeV. In Ref.~\cite{Dong:2022cuw} the authors explained the $\eta_1(1855)$ as the $K \bar K_1(1400)$ hadronic molecule, and this partial decay width was estimated to be only around 0.9~MeV. This significant difference may be useful when examining the nature of the $\eta_1(1855)$.

Based on the interpretation of the $\eta_1(1855)$ as the $\bar s s g$ hybrid meson of $I^GJ^{PC}=0^+1^{-+}$, we have also investigated its partner state, that is the $\bar q q g$ ($q=u/d$) hybrid meson of $I^GJ^{PC}=0^+1^{-+}$. Assuming its mass to be around $1600\pm100$~MeV, we calculate its partial decay widths into the $\eta \eta^\prime$ and $a_1(1260) \pi$ channels. The obtained results are summarized in Table~\ref{tab:result}, and this is our first scheme, labelled ``Scheme-I''.

Besides, we have considered another scheme, labelled ``Scheme-II'', where the $\eta_1(1855)$ is interpreted as the $\bar q q g$ hybrid meson of $I^GJ^{PC}=0^+1^{-+}$. In this case its partner state is the $\bar s s g$ hybrid meson of $I^GJ^{PC}=0^+1^{-+}$. Assuming their masses to be around $1855 \pm 9 ^{+6}_{-1}$~MeV and $2100\pm100$~MeV, we calculate their partial decay widths into the $\eta \eta^\prime$, $a_1(1260) \pi$, $f_1(1285) \eta$, $f_1(1420) \eta$, $K^*(892) \bar K$, $K_1(1270)\bar K$, and $K_1(1400)\bar K$ channels. During the calculations we have used $f_{a_1(1260)} = 238 \pm 10$~MeV~\cite{Yang:2007zt}. The obtained results are summarized in Table~\ref{tab:result}. It seems also possible to interpret the $\eta_1(1855)$ as the $\bar q q g$ ($q=u/d$) hybrid meson of $I^GJ^{PC}=0^+1^{-+}$, but this explanation is less favored considering: a) its total width is already larger than 250~MeV and there still exist some other possible decay channels not investigated, and b) the $\pi_1(1600)$ is a good isovector hybrid candidate whose isoscalar partner may have a similar mass.

To differentiate the above two schemes, and more important, to verify whether the $\eta_1(1855)$ is a hybrid meson or not, we propose to further search the other decay modes and its partner states in future BESIII, Belle-II, GlueX, LHC, and PANDA experiments.

%
\section*{Acknowledgments}
%

This project is supported by
the National Natural Science Foundation of China under Grants No.~12075019, No.~11975033, and No.~12070131001,
the Jiangsu Provincial Double-Innovation Program under Grant No.~JSSCRC2021488,
and
the Fundamental Research Funds for the Central Universities.

\end{document}